\documentclass{aa}  
\usepackage{graphicx}
\usepackage{txfonts}
\usepackage{natbib}
\bibpunct[; ]{(}{)}{,}{a}{}{,}
\def\asec{\ifmmode ^{\prime\prime}\else$^{\prime\prime}$\fi}

\def\it{\sl}
\def\degs{\ifmmode ^{\circ}\else$^{\circ}$\fi}
\def\amin{\ifmmode ^{\prime}\else$^{\prime}$\fi}
\def\asec{\ifmmode ^{\prime\prime}\else$^{\prime\prime}$\fi}
\def\fm{\hbox{$.\!\!^{\rm m}$}}            
\def\fdg{\hbox{$.\!\!^\circ$}}          
\def\farcs{\hbox{$.\!\!^{\prime\prime}$}}  

\def\psr{PSR~J1048$-$5832}
\def\degs{\ifmmode ^{\circ}\else$^{\circ}$\fi}
\def\amin{\ifmmode ^{\prime}\else$^{\prime}$\fi}
\def\farcm{\hbox{$.\mkern-4mu^\prime$}}
\def\eqalign#1{\null\,\vcenter{\openup1\jot \m@th
   \ialign{\strut\hfil$\displaystyle{##}$&$\displaystyle{{}##}$\hfil
   \crcr#1\crcr}}\,}
\begin{document}
   \title{   
Deep optical imaging of the $\gamma$-ray pulsar J1048$-$5832 with the VLT       
\thanks{Based 
on observations made with ESO telescope at the Paranal Observatory under
Programs 384.D-0386(A) and 386.D-0585(A).
}}

\author{A.~Danilenko\inst{1} 
\and A.~Kirichenko\inst{1,2}
\and J.~Sollerman\inst{3}
\and Yu.~Shibanov\inst{1,2}
\and D.~Zyuzin\inst{1,2}
}


\institute{
Ioffe Physical Technical Institute, Politekhnicheskaya 26,
St. Petersburg, 194021, Russia \\ 
danila@astro.ioffe.ru, shib@astro.ioffe.ru   
\and
St. Petersburg State Polytechnical Univ., Politekhnicheskaya 29, 
St. Petersburg, 195251, Russia \\ 
aida.taylor@gmail.com, da.zyuzin@gmail.com
\and
The Oskar Klein Centre, Department of Astronomy, Stockholm University,
AlbaNova, 106 91 Stockholm, Sweden \\
}


 
  \abstract 
    {\psr~is a 
    young radio-pulsar
    that  was recently detected in 
    $\gamma$-rays with \textit{Fermi},   
    and also in X-rays with \textit{Chandra} and  
    \textit{XMM-Newton}.  
    It powers a compact pulsar wind nebula visible in
    X-rays and  
    is in many ways similar to the Vela pulsar.}
    {We present deep optical 
    observations with the ESO Very Large Telescope 
    to search for optical counterparts of the pulsar and its nebula and 
    to explore their multi-wavelength emission properties.  
    }
    {The data were obtained in  
    $V$ and $R$ 
    bands and compared  with archival   
    data in other spectral domains.}
    {We do not detect the pulsar in the
    optical and derive informative 
    upper limits of $R$~$\gtrsim$ 28\fm1 and $V$~$\gtrsim$ 28\fm4
    for its brightness.  Using a
    red-clump star method, we estimate an interstellar extinction 
    towards the pulsar of $A_V$~$\approx$ 2 mag, 
    which is consistent with 
    the absorbing column density derived 
    form X-rays. 
    The respective distance agrees with the dispersion 
    measure distance.
    We reanalyse the \textit{Chandra} X-ray data  
    and compare the dereddened upper limits with the unabsorbed X-ray
    spectrum of the pulsar.  
    We find that regarding its optical-X-ray spectral properties 
    this $\gamma$-ray  pulsar is not distinct from
    other pulsars detected in both ranges. 
    However, like the Vela pulsar, it is very inefficient in the optical 
    and X-rays. 
    Among a dozen optical sources overlapping with the
    pulsar X-ray nebula  
    we find one with $V$~$\approx$ 26\fm9 and $R$~$\approx$ 26\fm3,  
    whose colour is 
    slightly bluer  
    then that of the field stars and  consistent with the peculiar  
    colours typical for pulsar nebula 
    features. It positionally coincides 
    with a relatively bright feature of the pulsar X-ray nebula, resembling 
    the Crab wisp and locating in 
    $\sim$~2\asec~from the pulsar.  We
    suggest this source as a  counterpart
    candidate to the feature.  
    }
%
    {Based on the substantial interstellar
    extinction towards the pulsar and its optical inefficiency, 
    further optical  studies 
    should be
    carried out at longer wavelengths.}      

\authorrunning{A.~Danilenko et al.}
\titlerunning{Deep optical imaging of \psr }
   \maketitle
%
\section{Introduction}
\label{sec1}
Rotation-powered pulsars are believed to be the most numerous of all 
$\gamma$-ray sources in the Galaxy.
Nevertheless, only about ten pulsars were discovered in $\gamma$-rays until 
the launch of the \textit{Fermi Gamma-ray Space Telescope} 
\citep{thompson2008RPPh}.
The number of $\gamma$-ray 
pulsars has now become about ten times larger, 
including many 
which have not yet been identified in 
the radio range\footnote{see https://confluence.slac.stanford.edu/display/GLAMCOG/ Public+List+of+LAT-Detected+Gamma-Ray+Pulsars}.  
The majority of the early known $\gamma$-ray pulsars are also identified 
in the optical and X-rays. This provides a unique possibility to compile 
multi-wavelength spectra and light curves 
for these objects for the study of the not yet clearly understood 
radiative mechanisms responsible for the pulsar emission.   
Only six such multi-wavelength objects are detected in 
the optical.
Increasing the number of optically identified  $\gamma$-ray pulsars is  
highly desirable and the \textit{Fermi} discoveries open up 
a new window for that. 

The first \textit{Fermi} catalogue \citep{abdo2010ApJS_Fermi_Catalog}
provided poor accuracy
(several arc-minutes) 
of the source spatial localisation. 
Therefore only those new $\gamma$-ray pulsars 
whose coordinates were known with higher accuracy from radio and/or X-ray 
observations were suitable  
for optical counterparts searches. 
Here we present  optical followup observations of one 
of the "radio-selected"  \textit{Fermi}-pulsars, \psr~(B1046$-$58),  
obtained with the  ESO Very Large Telescope (VLT).
This is a 124 ms Vela-like radio-pulsar with a characteristic age of 
20.4 kyr and a spin-down luminosity of 2~$\times$~10$^{36}$ erg s$^{-1}$, 
discovered in the radio by \citet{johnston1992MNRAS} and later in 
X-rays by \citet{gonzalez2006ApJ}. 
It was mentioned as a low significance $\gamma$-ray pulsar in the Third EGRET  
catalogue \citep{thompson2008RPPh} but was confirmed  
with \textit{Fermi} \citep{abdo2010ApJS_Fermi_Catalog}. 
The dispersion measure (DM) of 129 pc cm$^{-3}$ 
implies a distance $\approx$~2.7 kpc. A faint, presumably tail-like pulsar wind 
nebula (PWN) was also detected around 
the pulsar in X-rays \citep{gonzalez2006ApJ}. 

We imaged the pulsar  
field in $V$ and $R$ bands. 
\citet{mignani2011A&A} 
reported their 
non-detection of an optical counterpart based on our $V$-band data only. 
Here we analyse the data in both bands 
and examine colours of optical sources located close 
to the pulsar and within its PWN extent. 
We confirm the non-detection in the $V$ band and  
derive deep optical flux upper
limits for the pulsar in both bands within its 1$\sigma$
position uncertainty in X-rays. 
The observations and data reduction are described  
in Sect.~\ref{sec2}, our results are presented 
in Sect.~\ref{sec3} and discussed in Sect.~\ref{sec4}.  
\section{Observations and data reduction}  
\label{sec2}
\begin{table}[t]
\caption{Log of the VLT/FORS2 observations of 
\object{\psr}. }
\begin{center}
\begin{tabular}{lllll}
\hline\hline
   Date           &  Band         & Exposure               &  Airmass      & Seeing          \\
                  &               & [s]                    &               & [arcsec]      \\
\hline \hline     
   2010-01-09     &  $V$          &  15$\times$3           & 1.52          &  0.6--0.9      \\         
   2010-01-11     &  $V$          & 750$\times$4           & 1.22          &  0.5--0.7      \\
   2010-01-13     &  $V$          & 750$\times$8           & 1.24          &  0.4--0.6      \\                              
   2010-01-23     &  $V$          & 750$\times$4           & 1.22          &  0.5--1.0      \\
   2010-01-24     &  $V$          & 750$\times$8           & 1.21          &  0.4--0.7      \\
   2010-02-10     &  $V$          & 750$\times$8           & 1.23          &  0.5--0.8      \\
   2010-12-04     &  $R$          & 600$\times$5           & 1.42          &  0.6--0.7      \\
   2011-01-01     &  $R$          & 600$\times$13          & 1.47          &  0.5--0.9      \\
   2011-01-04     &  $R$          & 600$\times$5           & 1.42          &  0.5--0.6      \\
   2011-01-05     &  $R$          & 600$\times$5           & 1.26          &  0.4--0.6      \\
\hline                                                                                 
\end{tabular}
\end{center}
\label{t:log}
\end{table}

\subsection{Observations}
\label{obs}
The pulsar field was imaged in the $V_{HIGH}$ and $R_{SPECIAL}$ bands 
with the FOcal Reducer and low dispersion Spectrograph 
(FORS2\footnote{For instrument details see {http://www.eso.org/instruments/fors/}}) 
at the VLT/UT1 (ANTU)     
during several service mode runs 
in 2010,  and 2011 (see Table~\ref{t:log}).
The observations were performed with the Standard Resolution collimator  
providing a pixel size of 0\farcs25 
(2$\times$2 binning) and field size of 6\farcm8~$\times$~6\farcm8.  
Sets of twelve- and ten-minute dithered exposures were obtained in the $V$ and
$R$ bands, respectively. Three short, 15 sec, 
exposures were taken in the $V$ band to minimise 
the number of saturated sources in the crowded pulsar field and were used  
for astrometry.  The observing conditions were 
photometric during the runs, 
with seeing varying from 0\farcs4 to 1\farcs0.

Standard data reduction, including  bias subtraction, flat-fielding, 
cosmic-ray removal, and geometric distortion corrections,
was performed making use of the {\tt IRAF} and {\tt MIDAS}  tools.      
We then aligned and combined all 
individual frames in each of the bands, using a set of unsaturated stars. 
The alignment accuracy was $\la$~0.1 of a pixel.
The resulting mean seeing values were 0\farcs63 and 0\farcs64, 
and the integration times were  
24 and 16.8 ks, for the combined $V$ and $R$ images, respectively.  

\subsection{Astrometric referencing} 
\label{astroref}
For astrometric referencing the shallow $V$ band 
(Table \ref{t:log})  images were used. 
The positions of the astrometric standards  
from the USNO-B1 astrometric catalogue\footnote{see 
http://www.nofs.navy.mil/data/fchpix/}
were used as a reference.  
To minimise uncertainties 
caused by overlapping stellar profiles in the crowded FOV, 
we selected only 15 isolated non-saturated stars. 
Their pixel coordinates 
were derived using the {\tt IRAF} task {\sl imcenter} with 
an accuracy of $\la$~0.007 pixels. 
The {\tt IRAF} task {\sl ccmap},
allowing for the image scaling, shift,
and rotation, was  
applied for the astrometric transformation of the image. 
Formal {\sl rms} uncertainties of the  
astrometric fit were $\Delta$RA~$\la$ 0\farcs056 
and $\Delta$Dec~$\la$ 0\farcs060, and  the fit residuals 
were $\la$~0\farcs17, 
consistent with the nominal catalogue uncertainty of $\approx$~0\farcs2. 
The combined deep $V$ and $R$ images were 
aligned to the short $V$ reference frame with an accuracy of $\la$~0\farcs01.   
The resulting conservative 1$\sigma$ referencing uncertainty 
for the combined  images is   
$\la$~0\farcs2 in both RA and Dec.

In order to compare optical and X-ray data, 
we also performed astrometric referencing of the best spatial resolution  
archival X-ray image of \psr~obtained with \textit{Chandra}/ACIS\footnote{Obs. ID 3842, date 2003.10.08, 
Exp. time 36 ks, PI V. Kaspi.}. 
In the  exposure-corrected ACIS-S3 chip image, where the pulsar is located,   
we found a dozen of point-like objects detected 
at $\ga$~3$\sigma$ significance.  
We identified them with relatively bright optical 
reference objects  from the  USNO-B1 catalogue. 
Their image positions were defined
using the {\tt CIAO} {\it celldetect} tool with 
an  accuracy of  0.5--3.0 of the ACIS  pixel size ($\approx$~0\farcs5). 
Resulting formal {\sl rms} uncertainties of the   astrometric fit  
were $\Delta$RA~$\approx$ 0\farcs424 and 
$\Delta$Dec~$\approx$ 0\farcs22 with maximal 
residuals $\la$~0\farcs83 and $\la$~0\farcs54 
in RA and Dec, respectively. 
Combining the latter  with the catalogue uncertainties, 
conservative 1$\sigma$ X-ray image astrometric uncertainties are 
$\Delta$RA~$\la$ 0\farcs85 and
$\Delta$Dec~$\la$ 0\farcs58. 
The shift between 
the raw and transformed images was
insignificant, $\sim$~0\farcs1.   

Using {\sl celldetect} we obtained the coordinates of the point-like pulsar 
X-ray counterpart, RA~= 10:48:12.640 and Dec~= $-$58:32:03.50, which are 
compatible with those reported by \citet{gonzalez2006ApJ}. 
The radii and positional angle of the 1$\sigma$-error ellipse of the source 
position are  0\farcs335, 0\farcs266, 168\fdg663, respectively. 
Accounting for the  image referencing uncertainties we estimate a conservative 
pulsar X-ray coordinate errors of 0\farcs92 and 0\farcs64 in RA and Dec, 
respectively. 
Combining them with the optical
referencing uncertainty
we obtained the RA and Dec radii of 
the 1$\sigma$-error ellipse of the pulsar X-ray 
position on the optical images as
0\farcs94 and 0\farcs67,
respectively.         

\subsection{Photometric calibration}                
\label{photcal}
The photometric calibration was carried out using 
standard stars from the  photometric standard fields 
\object{E3}, \object{NGC2298}, \object{NGC2437}, 
and \object{PG1525} \citep{stetson2000PASP} observed during the same nights as
the target. We fixed the atmospheric extinction coefficients
at their mean values adopted from the VLT home page: 
k$_V$~= 0\fm14~$\pm$~0\fm01 and k$_R$~= 0\fm09~$\pm$~0\fm01.  
The resulting magnitude zero-points for
the combined images 
were
$V^{ZP}$~= 27\fm97~$\pm$~0\fm02 and
$R^{ZP}$~= 28\fm12~$\pm$~0\fm02,
and  colour-term 
coefficients  in respective photometric equations\footnote{ see e.g. ``A User's Guide to Stellar CCD Photometry with IRAF'' by P.~Massey and L.~Davis,
http://iraf.net/irafdocs/} 
are  0.15~$\pm$~0.03 
and $-$0.01~$\pm$~0.03. 
The errors include 
the statistical measurement and extinction coefficient uncertainties, 
and marginal  variations from night to night. 
\begin{figure*}[t]
  \setlength{\unitlength}{1mm}
  \begin{center}
    \begin{picture}(145,59)(0,0)
      \put (-12,0) {\includegraphics[scale=0.43]{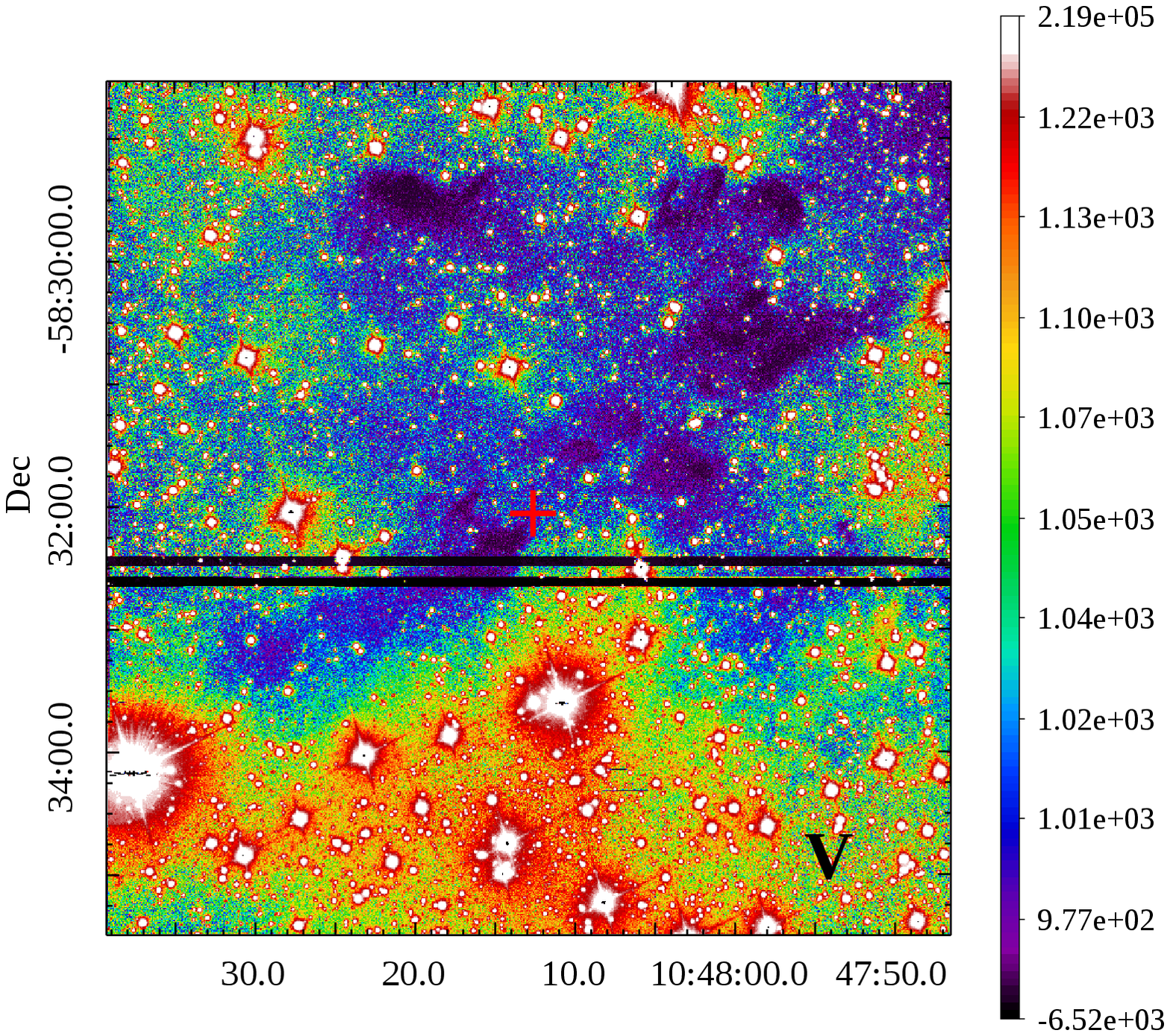}}
      \put (80,0) {\includegraphics[scale=0.43]{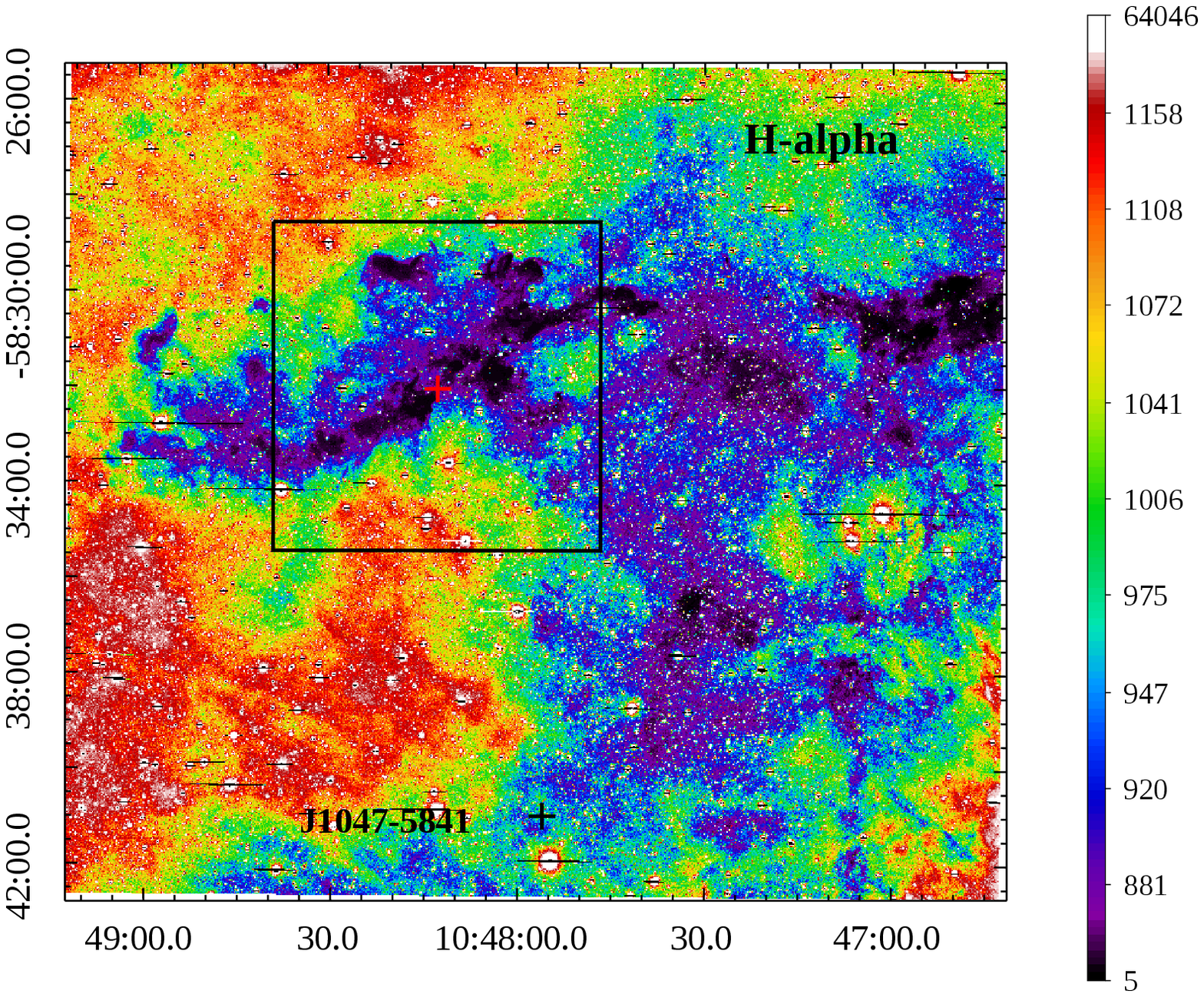}} 
    \end{picture}
  \end{center}
  \caption{The field of \psr~obtained in the
  $V$ band with the VLT ({\sl left}),
  and in H$\alpha$  with the  CTIO ({\sl right}). 
  The dark horizontal lines 
  in the VLT image are due to a gap between the two FORS2/CCD chips and dithered  exposures.   
  The box in the H$\alpha$ image shows the VLT FOV.
  The  red and black crosses are the
  positions of  J1048$-$5832 and a nearby neutron star  
  RRAT J1047$-$5841.
  }
  \label{fig:1}
\end{figure*}

\section{Results}
\label{sec3}
\subsection{Overview of the pulsar field} 
\label{sec:overview} 
The $\sim$6\amin$\times$6\amin~VLT FOV
in the $V$ band ({\sl left} panel of Fig.~\ref{fig:1})
shows a complicated structure of the pulsar
environment with a dark feature extended over the entire field.  
This is also seen  in
the $R$ band and is fully consistent with a large scale  structure  in the H$\alpha$ 
archival image\footnote{Obtained with the CTIO 4-m 
telescope as a part of ChaMPlane survey 
\citep{grindlay2005ApJ}} ({\sl right} panel of Fig.~\ref{fig:1}). 
The pulsar is  near the middle of
its eastern part. 
In the \textit{Spitzer} archival 
images  at 8
and 24 $\mu$m\footnote{GLIMPSE    
project, PI S. Majewski.}  the dark
part  is filled  with bright infrared emission,  
likely a signature of
a warm dust, which is typical for  star
forming regions.
Another neutron star, RRAT
\object{J1047$-$58} is located  
$\sim$~15\amin~from the pulsar. 
Its distance     
$\approx$~2.33 kpc 
\citep{keane2011MNRAS} is similar to that of J1048$-$5832.  
Both objects are only 1\fdg5 
northwards of the centre of the
Carina complex, one of the largest
H~II  regions in our Galaxy
at a distance of $\sim$~2.3 kpc \citep{smith2006ApJ}.  
It contains a neutron star, 2XMM J104608.7$-$594306 \citep{pires2012A&A},
and 6 potential  neutron stars candidates 
\citep{townsley2011ApJS} showing
past supernova activity. 
The complicated J1048$-$5832 environment 
is thus likely linked  to the north edge  of    
the Carina complex. 
\begin{figure*}[tbh]
  \setlength{\unitlength}{1mm}
  \begin{center} 
    \begin{picture}(160,110)(0,0) 
      \put (-5,51)   {\includegraphics[scale=0.33,clip]{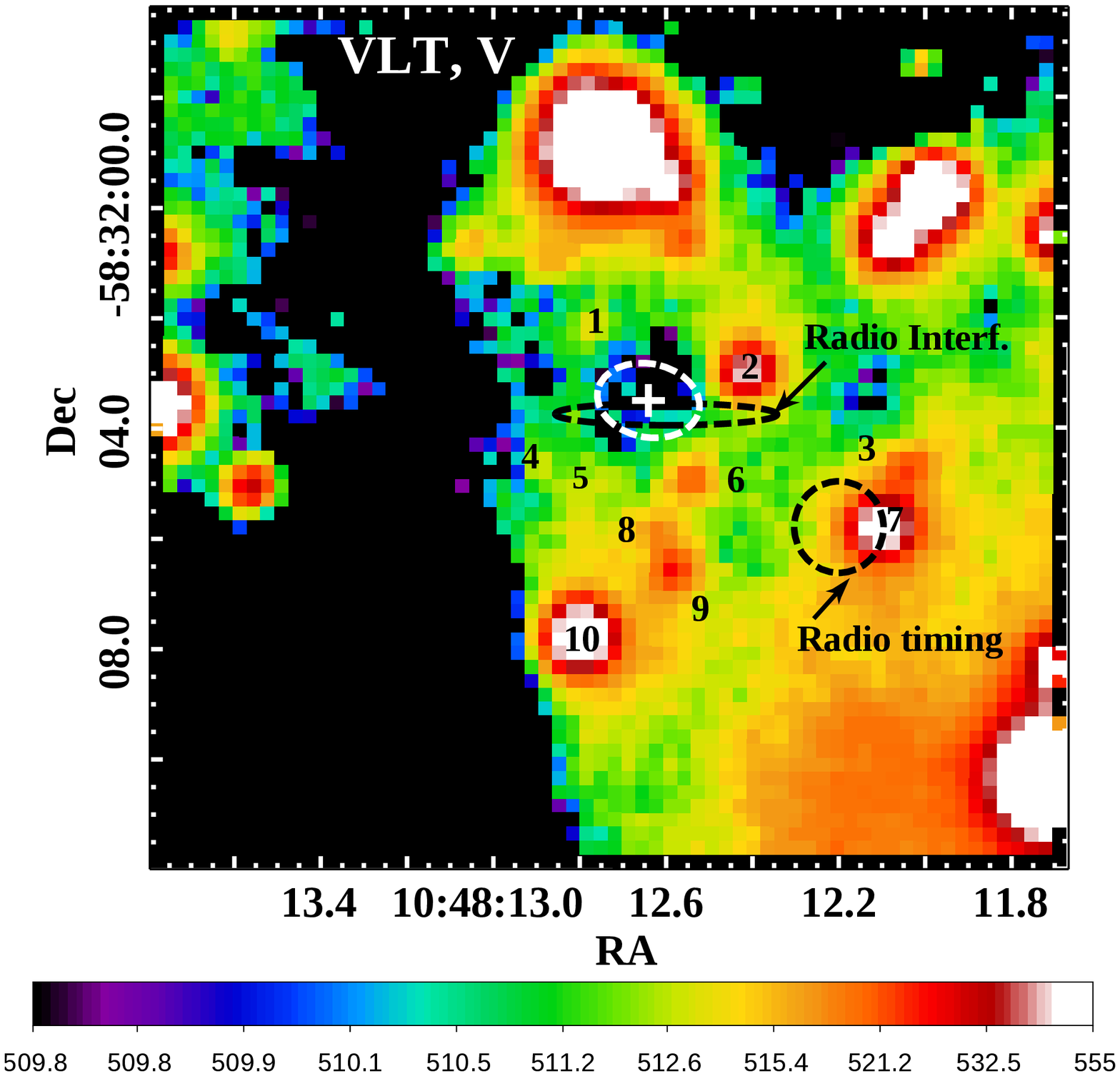}}
      \put (60,50.5) {\includegraphics[scale=0.32,clip]{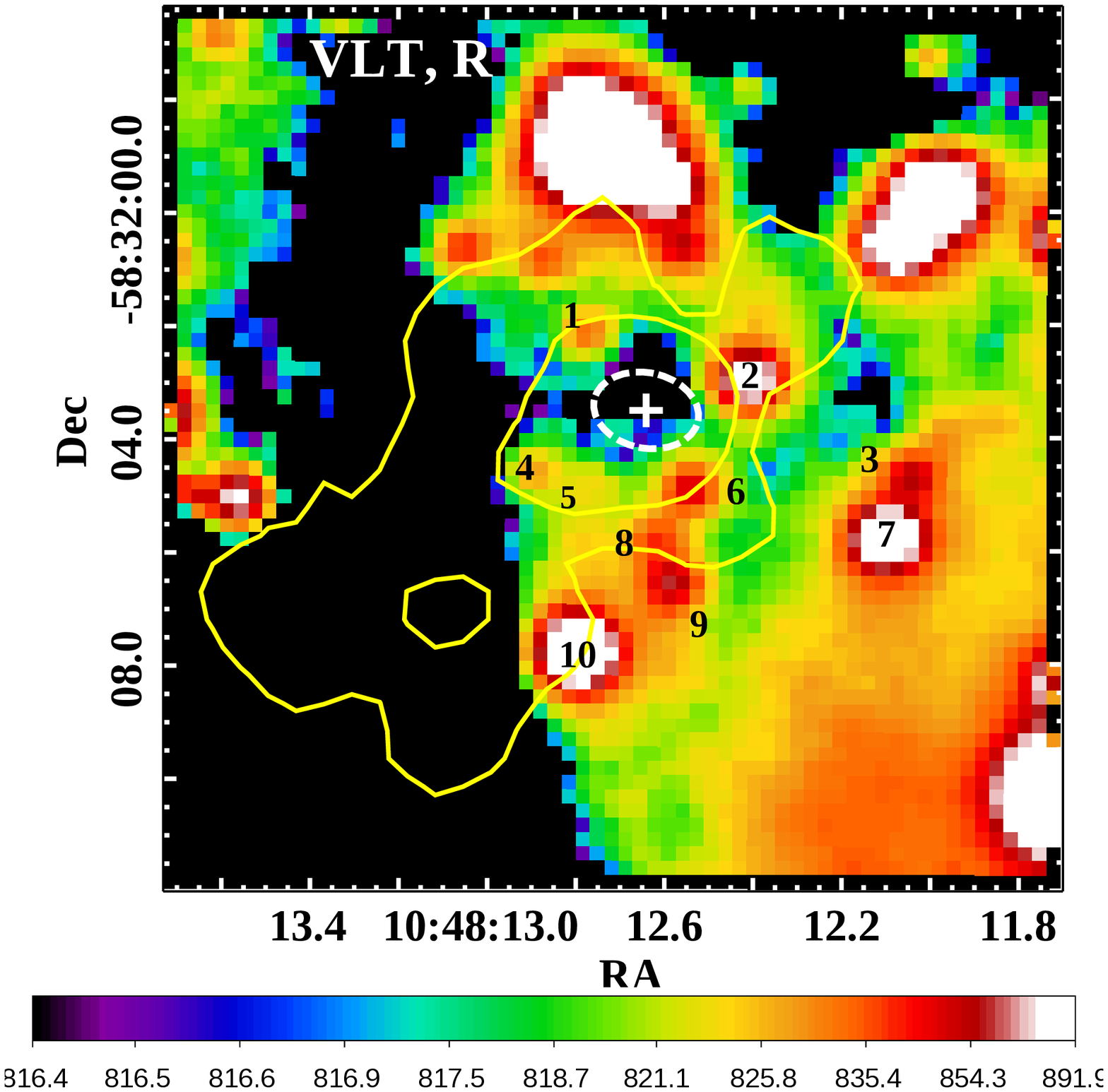}}
      \put (129,48)  {\includegraphics[scale=0.215,clip]{V_vs_V-R.eps}}
      \put (-3,-10)  {\includegraphics[scale=0.33,clip]{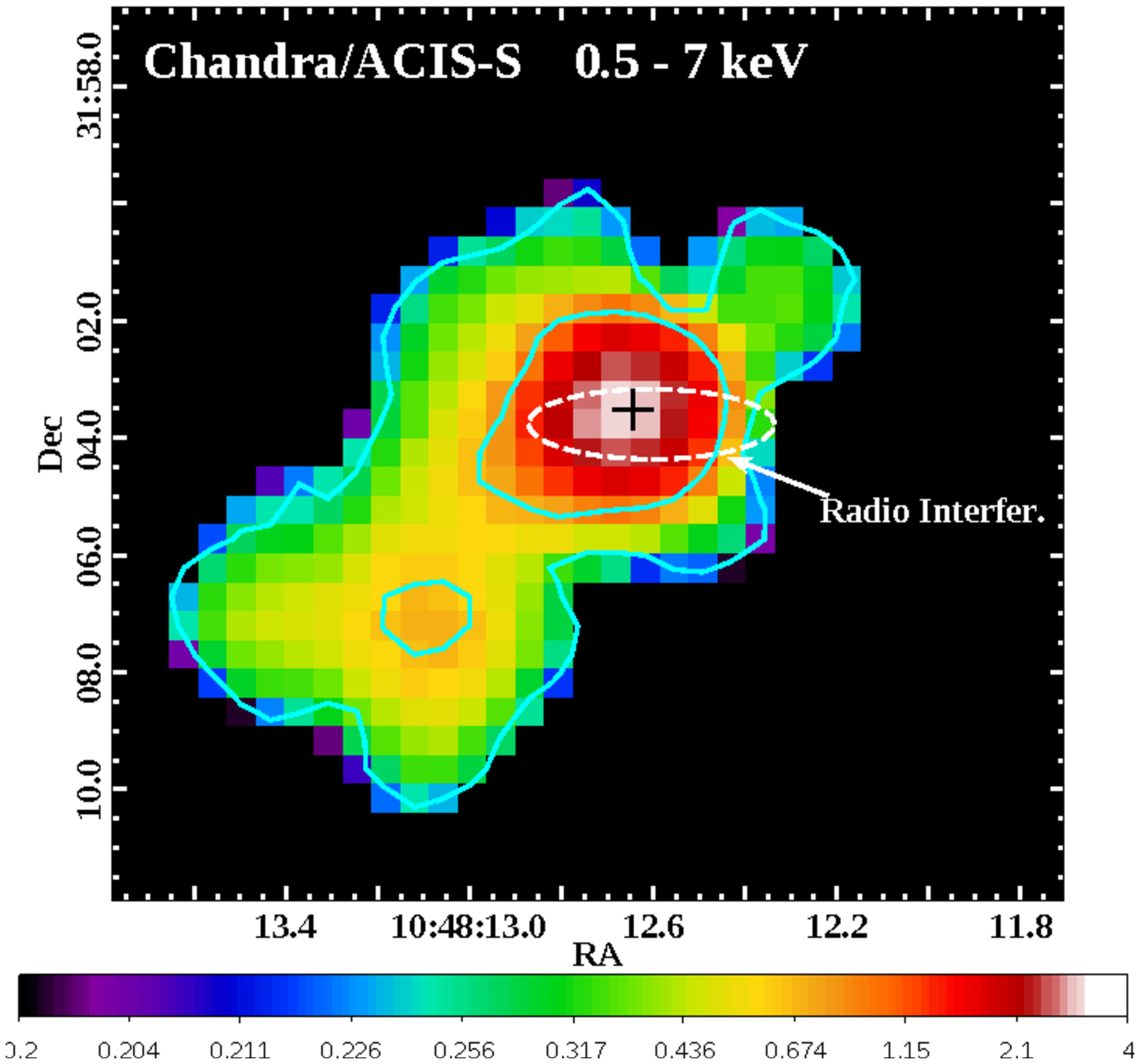}}    
    \end{picture} 
  \end{center}
  \vspace{-47mm}
  \hfill\parbox[b]{97mm}{    
  \caption{The $\sim$~14\farcs~$\times$~14\farcs~fragment of 
  the VLT $V$ ({\sl top-left}) and $R$ ({\sl top-middle}) band  optical,   
  and \textit{Chandra} 0.5--7 keV X-ray ({\sl left-bottom}) images 
  of the \psr~vicinity. The
  \textit{Chandra} image is smoothed with a
  four pixel Gaussian  kernel.  
  Colour-bars  show  brightness scales in 1000 counts in the optical  
  and in counts in X-rays.  
  The cross marks the X-ray position 
  of the pulsar.  White-dashed,  and black-dashed  
  ellipses show its 1$\sigma$ uncertainties
  in X-rays and two radio  observations, respectively.
  Yellow contours in the $R$-band image are overlaid 
  from the X-ray image to indicate the PWN structure.
  Optical sources overlapping 
  with the error ellipses and 
  X-ray PWN are labelled by numbers.
  {\sl Top-right:} The observed colour-magnitude diagram of \psr~field stars.
  The stars labelled in the  images are highlighted.  
  }
  \label{fig:pulsar_vicinity}
  } 
\end{figure*}

\subsection{Examining the pulsar vicinity} 
%
In Fig.~\ref{fig:pulsar_vicinity} we compare images of the pulsar 
vicinity in the $VR$ bands and X-rays.  
The X-ray image  is corrected 
for the ACIS exposure map and smoothed with a four pixel Gaussian kernel 
to  better show the PWN shape. The X-ray position of 
the pulsar and its 1$\sigma$ uncertainty are marked 
together with the radio interferometric 
\citep{stappers1999MNRAS}\footnote{Instrument ATCA, 
date 13-05-1997, epoch 50581, 
RA~= 10:48:12.6(3), Dec~=$-$58:32:03.75(1) J2000.} and 
timing \citep{wang2000MNRAS}\footnote{Instrument Parkes, date 25-02-1993--29-03-1997, 
epoch 49043--50536,  RA~= 10:48:12.2(1),
Dec~= $-$58:32:05.8(8) J2000.} 
ones. The X-ray position is in a good agreement with
the interferometric one.  This fact  and our accurate X-ray astrometric 
referencing allow us to
use the X-ray  position  
as a reliable reference point in searching for 
the pulsar  counterpart.
The timing position  is shifted
significantly and likely suffers from  
systematic errors.    
The contours  of the brightest regions and the outer 
boundary of the X-ray PWN are overlaid on the $R$ image.  

We do not resolve any reliable optical source within the pulsar 
1$\sigma$ X-ray error ellipse, but   
there are several 
optical sources overlapping with 
the PWN and radio ellipses.
They are labelled by numbers and  
can be considered as potential counterparts of the pulsar,
if it has a high proper motion,
or its PWN structures.

\subsection{Photometry and colour-magnitude diagram}
Pulsars and PWNe 
typically have peculiar colours 
as compared to 
stars. To investigate whether the marked 
sources are associated with the pulsar and/or its PWN we  
performed Point Spread Function
(PSF) photometry
using the \texttt{psf} and \texttt{allstar} tasks of the IRAF \texttt{DAOPHOT} 
package \citep{stetson1987PASP}.
We set the psf-radius at ten pixels, where the bright 
isolated unsaturated stars selected for the PSF-construction 
merged with the background. The fit-radius and aperture radius 
for the PSF stars and the
preliminary aperture photometry of the other stars 
were 2.5 pixels, while  
the annulus/dannulus for local
background extractions were 15/10 pixels. 
We made aperture corrections based 
on photometry of bright unsaturated isolated field stars. 
\begin{table}[tbh]
  \caption{Magnitudes and fluxes of the 
  optical sources labelled in
  Fig.~\ref{fig:pulsar_vicinity} and  
  \psr~upper limits (psr).
  Numbers in brackets are
  1$\sigma$ uncertainties referring to the last significant digits quoted. 
  Stars 6 and 7 were labelled by $D$ and $C$ respectively in \citet{mignani2011A&A}}
  \begin{tabular}{ccccc}
    \hline \hline
    Star    & $V$ (mag)       & $R$ (mag)  & $\log$ F$_{V}$ ($\mu$Jy) & $\log$ F$_{R}$ ($\mu$Jy) \\
    \hline 
    1       & 26.6(1)         & 25.22(6)   & $-$1.11(6)               & $-$0.61(2) \\
    2       & 24.62(3)        & 23.36(2)   & $-$0.29(1)               & 0.134(7) \\
    3       & 26.6(2)         & 24.88(8)   & $-$1.09(9)               & $-$0.47(3) \\
    4       & 26.9(2)         & 25.68(8)   & $-$1.22(7)               & $-$0.79(3) \\
    5       & 26.9(2)         & 26.3(2)    & $-$1.23(9)               & $-$1.06(7) \\
    6($D$)  & 26.1(1)         & 24.67(4)   & $-$0.90(4)               & $-$0.38(2) \\
    7($C$)  & 23.92(2)        & 22.77(1)   & $-$0.015(6)              & 0.368(4) \\
    8       & 26.5(1)         & 25.22(6)   & $-$1.08(5)               & $-$0.61(2) \\
    9       & 25.62(6)        & 24.31(4)   & $-$0.69(2)               & $-$0.24(2) \\
    10      & 23.41(2)        & 22.24(2)   & 0.189(8)                 & 0.585(8) \\
    psr     & $\ga$ 28.4      & $\ga$ 28.1 & $\la$ $-$1.81            & $\la$ $-$1.76
    \\
    \hline
  \end{tabular}
  \label{t:mag}
\end{table}
The derived $3\sigma$ detection-limits for a point-like object 
for a half-arcsecond aperture
centred at the  pulsar X-ray position 
are $V^{up}$~$\approx$ 28\fm4 and $R^{up}$~$\approx$ 28\fm1,
accounting for the aperture corrections. 
Magnitudes of the sources marked in
Fig.~\ref{fig:pulsar_vicinity} 
and the above  upper limits  are collected in Table~\ref{t:mag}
where  uncertainties include the  measurement 
and   calibration
errors. The  magnitudes were transformed into flux densities  
using zero-points from \citet{fukugita1995PASP}.

Comparing the $V$ magnitudes of 
stars 6 and 7 to those estimated by 
\citet[][stars $D$ and $C$ in their notations]{mignani2011A&A},   
we find  that for     
star 7 ($C$), overlapping with the pulsar timing position,  
their estimate, $\sim$~24\fm,  is 
consistent with ours,  
23\fm92~$\pm$ 0.02. 
However, their reported magnitude for
the fainter object 6 ($D$) is about 
0\fm5 fainter than our measurement.     

The output of the \texttt{allstar} task was also used for  photometry of 
field stars and  construction
of the colour-magnitude diagram
(Fig.~\ref{fig:pulsar_vicinity}). 
To exclude unresolved blends, partially resolved galaxies, saturated stars, 
and stars incorrectly
cross-identified in both bands, we 
selected only the stars with the \texttt{allstar} output parameters satisfying 
the following conditions: $\chi^2$~$\la$ 1.5;
sharpness $\la$~1;  
position differences in $V$ and $R$ bands $\la$~0.7 pixel. 
These criteria were also fulfilled for the sources 
listed in Table~\ref{t:mag}. 
The resulting sample contains about
3000  stars.
The objects with  magnitudes 
$\la$~20\fm5 are saturated and not included.   
The sources from the pulsar vicinity are
highlighted in the diagram 
and numbered as in the images.  
All of them,  
except possibly source 5, are well within the distribution formed by
the majority of field stars. They are likely belong to
the main-sequence branch and, thus, 
are unlikely to be associated with the pulsar. 

Within the uncertainties the colour  
$V-R$~= 0.6~$\pm$ 0.3  of source 5 is 
compatible  with the $V-R$~$\la$ 0.7 typical for  
pulsar/PWNe optical counterparts,   
which are usually detected as faint blue objects. 
For instance,  $V-R$ is 0.4 for the Crab pulsar \citep{percival1993ApJ} 
and 0.7 for its PWN knot \citep{sandberg2009A&A}.        
An association with the point-like pulsar is, however, 
excluded due to the large offset,  
1\farcs7~$\pm$~0\farcs7, from the  pulsar 
X-ray position.  
At $d$~= 2.7
kpc this would imply an unrealistically high pulsar transverse velocity  
of  3100~$\pm$ 1300 km~s$^{-1}$
on the 7 yr time-base between the \textit{Chandra} and VLT 
observations. The absence of any significant shift on the  6 yr time-base
between the interferometric and X-ray 
positions excludes such a motion. 
\begin{figure*}[tbh]
  \setlength{\unitlength}{1mm}
  \begin{picture}(160,113)(0,0) 
    \put (5,52) {\includegraphics[scale=0.38]{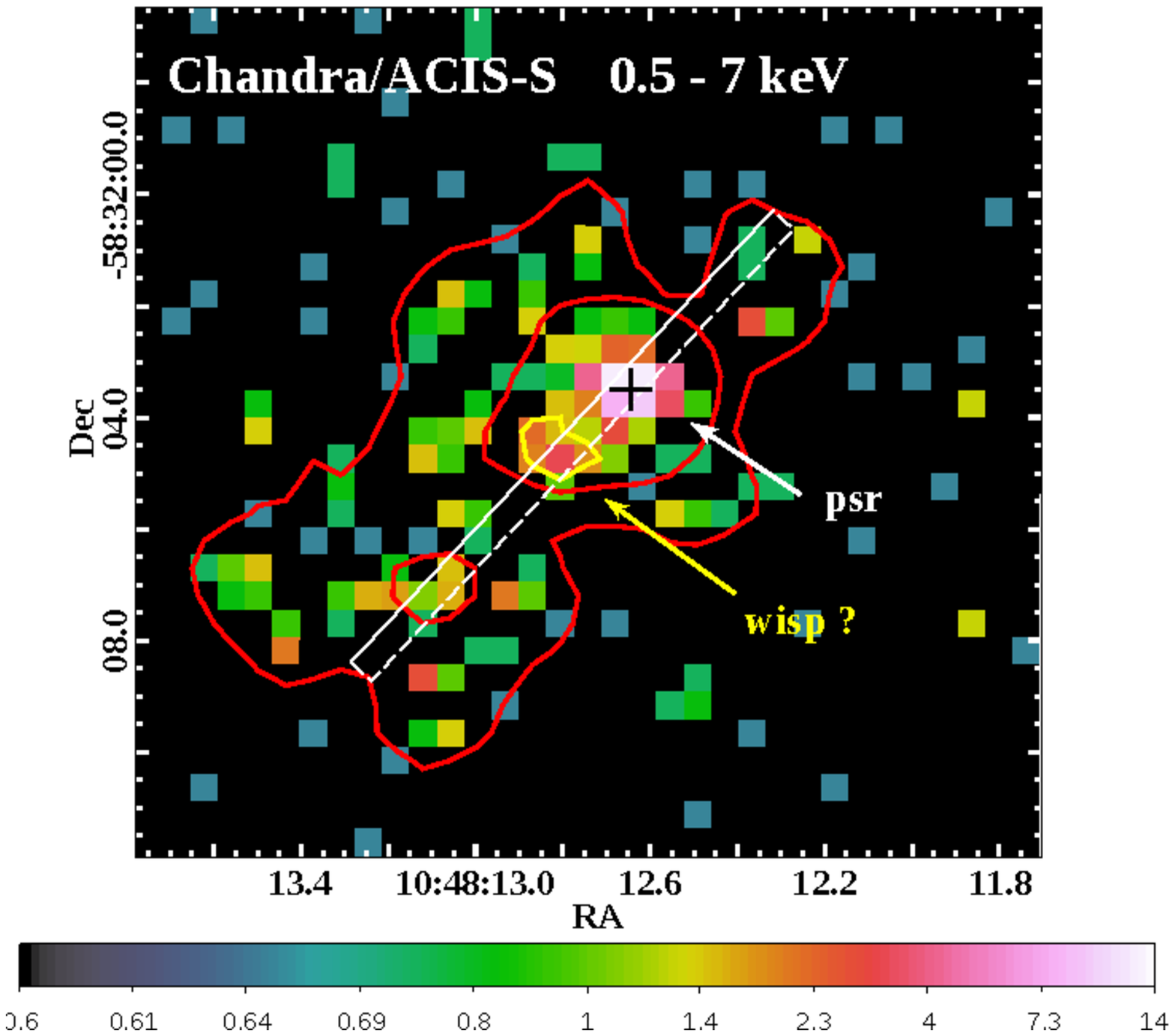}}
    \put (0,-10)  {\includegraphics[scale=0.39]{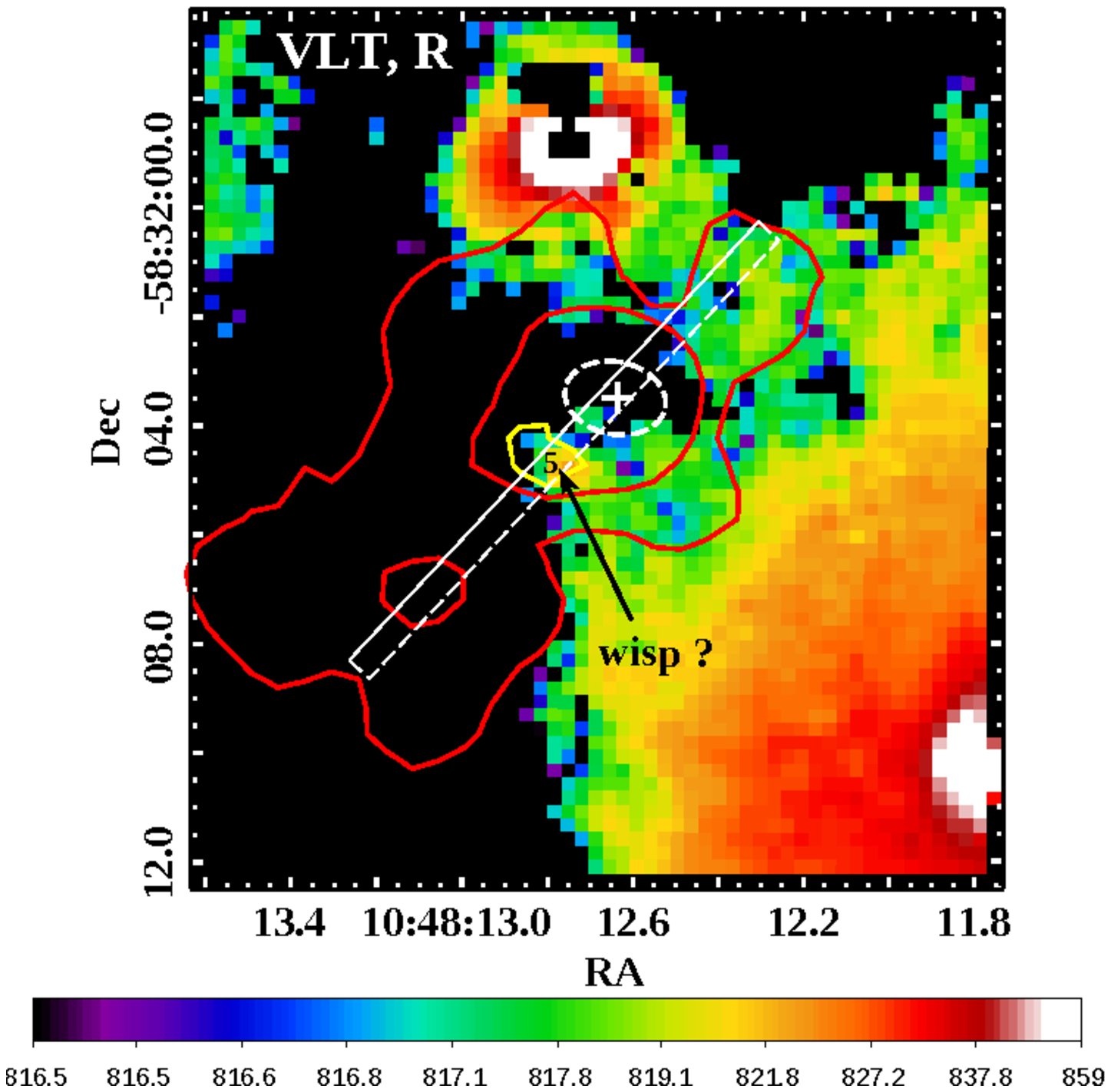}}
    \put (95,105) {\includegraphics[scale=0.28,angle=-90]{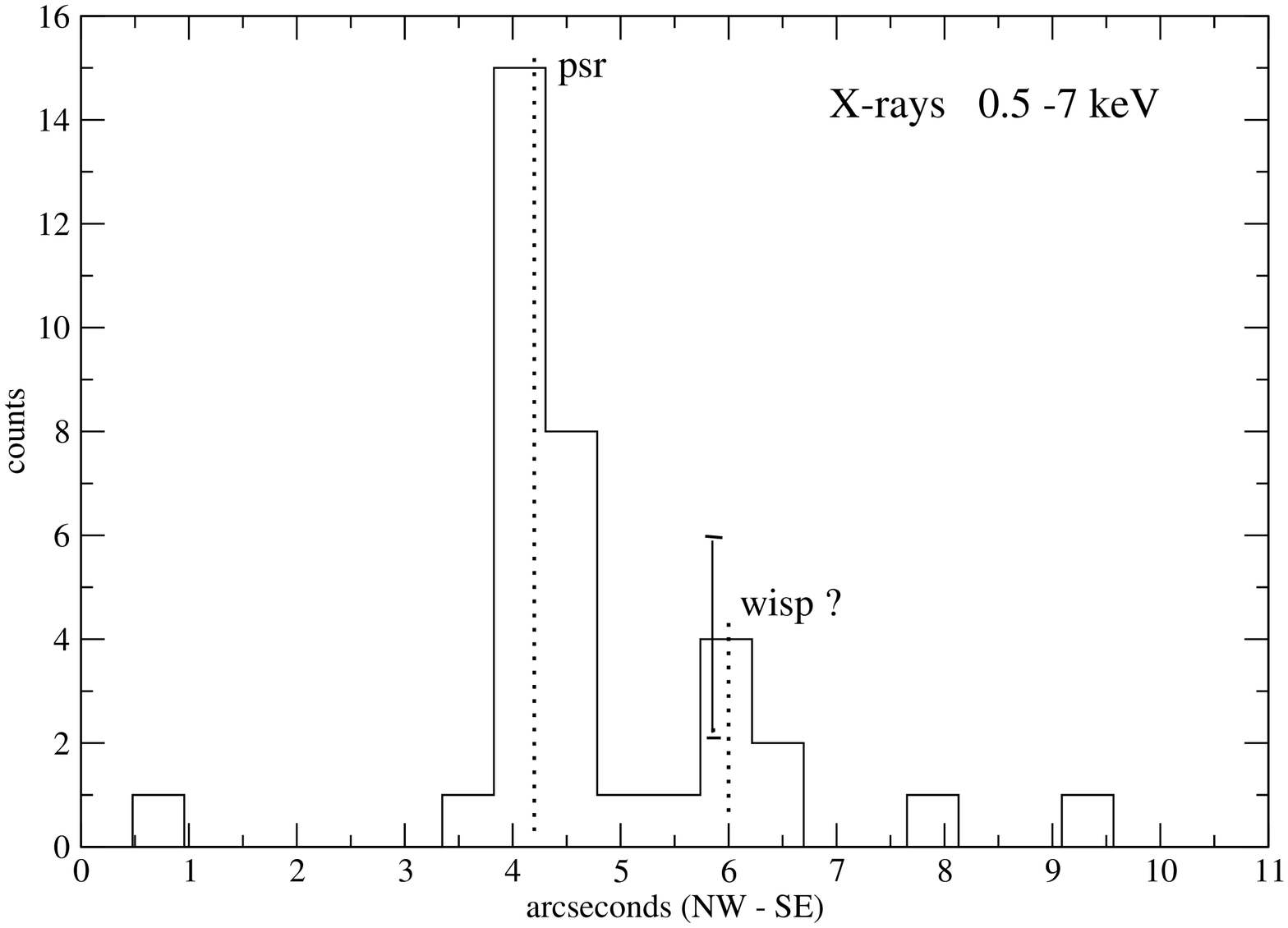}}
    \put (93.5,50) {\includegraphics[scale=0.28,angle=-90]{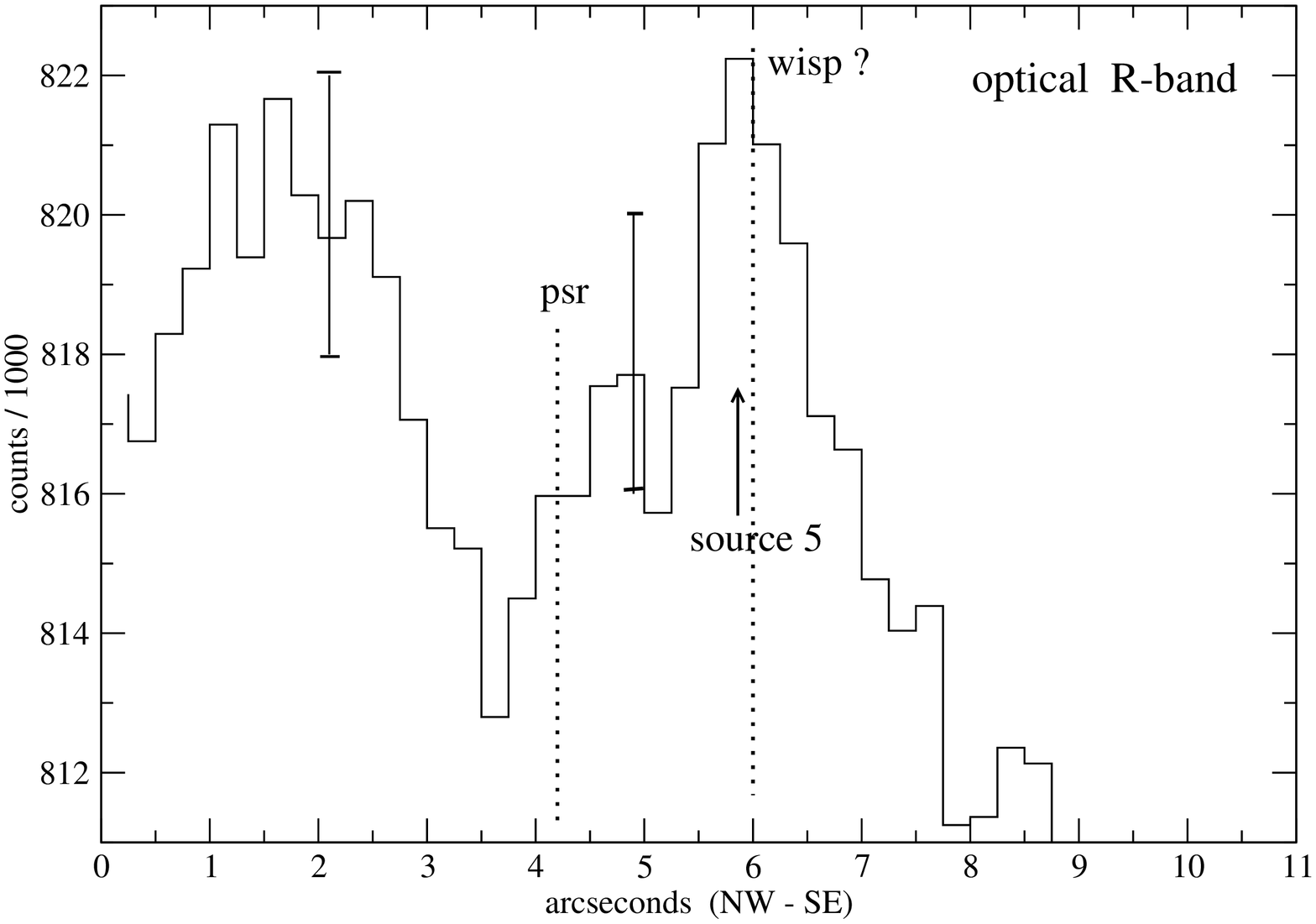}}
  \end{picture} 
  \caption{ The non-smoothed
  \textit{Chandra} image ({\sl top})
  and  star-subtracted VLT image  
  ({\sl bottom}).
  Magenta contours  are X-ray contours
  of the PWN from Fig.~\ref{fig:pulsar_vicinity}. 
  The cross  and dashed ellipse are  the position of the pulsar and its 
  1$\sigma$ uncertainty.  The
  yellow contour marks a relatively bright 
  X-ray structure of the PWN 
  2\asec~southeast of the pulsar, presumably a Crab-like wisp.  
  It spatially coincides with  the optical source 5, as   
  can also be seen from the spatial brightness profiles extracted 
  from a slice along the PWN major axis 
  (a white-dashed rectangular on both images) 
  presented in the {\sl left}
  panels, where  vertical dotted lines  indicate the X-ray positions 
  of the pulsar and the wisp.  
  Error-bars indicate typical brightness uncertainties.         
  }
  \label{fig:pulsar_vicinity2}
\end{figure*}

\subsection{Possible counterpart of the PWN}
In Fig.~\ref{fig:pulsar_vicinity2} we compare 
the non-smoothed  X-ray image of the pulsar vicinity 
with that in the $R$ band where  background stars have been subtracted 
except for source 5.   
There is a 4-$\sigma$ significance compact  
X-ray structure within the
PWN about 2\farcs0 
southeast of the pulsar. The structure
is outside the pulsar PSF FWHM and contains $\sim$~12--14 
source counts within the yellow region constraining its boundaries. 
It is reminiscent of the 
wisps in the Crab PWN observed in the optical and X-rays. 
The structure is labelled as ``wisp ?'' and yields an apparent drop-like shape 
of the brightest region around the pulsar in the smoothed X-ray image 
of Fig.~\ref{fig:pulsar_vicinity}. 
It spatially coincides with source 5. This is 
underlined by the yellow X-ray region overlaid on the optical image and by 
the spatial brightness profiles 
along the major PWN axis 
({\sl right} panels of Fig.~\ref{fig:pulsar_vicinity2}). 
The putative wisp can also be resolved in the X-ray profile 
published by \citet{gonzalez2006ApJ}, although
it is less pronounced.

The number density of
sources observed within the VLT FOV in the    
brightness range 26\fm9~$\la$ $V$
$\la$~28\fm4, consistent with the source  5  brightness,   
is $\sim$~0.008 objects~arcsec$^{-2}$. 
The respective confusion probability 
to find an unrelated point-like  optical source within 
the 90\% \textit{Chandra}  
positional uncertainty ellipse of
the putative wisp  is  $\sim$~2\%  and  it becomes  considerably 
smaller, $\sim$~0.2\%,  if we additionally constrain the colour
$(V-R)$~$\la$ 0.9, as is in the
source 5 case. 

There are no other reliable potential counterparts of the PWN in our star
subtracted images.  
A 1.5$\sigma$ $R$ flux enhancement
seen  within the pulsar X-ray error ellipse 
may indicate the presence of a
faint pulsar counterpart candidate,   
but is consistent  
with a background fluctuation and the pulsar
upper limits derived above.

\subsection{Re-examination of the X-ray spectra}
\label{sec:reexam}
It is useful to compare  the pulsar
optical upper limits    
with its X-ray spectral data. 
\citet{gonzalez2006ApJ}
reported  a spectral analysis  of  available 
\textit{Chandra} and \textit{XMM-Newton} 
data but only for the combined emission of the pulsar+PWN system. 
To examine the X-ray spectrum of the pulsar itself 
we performed an independent  
analysis. We first
reanalysed  the pulsar+PWN spectrum
using the data from both
instruments  
and our results are fully consistent with
the published ones.  
The spectrum is described  
by an absorbed power-law, whereas the
blackbody model 
gives an unrealistically high 
neutron star temperature.

\begin{figure}
  \begin{center}
    \includegraphics[scale=0.37,clip]{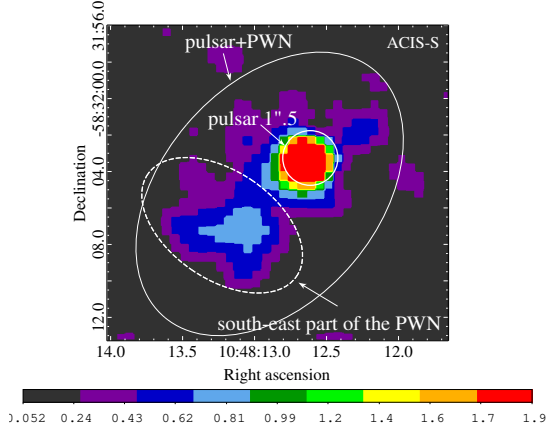}
  \end{center}
  \caption{The fragment of 
  the \textit{Chandra}/ACIS image
  of the \psr~field, where 
  ellipses mark the apertures
  used to extract spectra of the
  pulsar and PWN. 
  }
  \label{fig:acis-cont}
\end{figure}

We then focused on the \textit{Chandra}/ACIS data where 
the pulsar is  spatially
resolved from the PWN (Fig.~\ref{fig:pulsar_vicinity2}).     
We used three apertures  shown in Fig.~\ref{fig:acis-cont} 
to extract the spectra of the pulsar, the  pulsar+PWN system,  
and the south-east tail of the PWN. 
The numbers of source counts 
were 71~$\pm$ 9, representing    
$\ga$~80\% of the emission from 
the point-like pulsar,  176~$\pm$ 5, 
and 50~$\pm$ 9, respectively.  
The background was extracted from a
15\asec~circular aperture located
$\sim$40\asec~north-east of the pulsar 
in a region free from any sources.    
We fitted the absorbed power-law model to the extracted unbinned spectra  
in  the 0.5--10.0 keV range using
the \texttt{Xspec v.12.7.1} \citep{arnaud1996ASPC} and   
C-statistics \citep{cash1979ApJ,wachter1979ApJ}. 
Along with  C values the fit qualities were estimated using the 
\texttt{goodness task}\footnote{The \texttt{Xspec} \texttt{goodness} 
task simulates data with 
given response files and model. 
The fit is good when about 50\% of the simulated  
spectra have the value of C less than that of 
the data in question.}. The best fit parameters, C values,
and  energy bin numbers (nbins)  are presented in Table~\ref{t:x-fit}. 
The pulsar+PWN fit 
is in agreement with that  
obtained by \citet{gonzalez2006ApJ}.  
\begin{table}[b]
  \caption{The absorbing column density $N_H$, photon index $\Gamma$, and normalisation factor PL of absorbing power-laws describing    
  the \textit{Chandra} spectra (see Fig.~\ref{fig:acis-cont} and text). Errors correspond to $\Delta$C = 1. 
  }
  \begin{tabular}{cccc}
    \hline\hline
    $N_H$                  & $\Gamma$               &  PL$_{norm}$                                & C (nbins)\\ 
    10$^{22}$  cm$^{-2}$   &                        &  10$^{-6}$ ph cm$^{-2}$ s$^{-1}$ keV$^{-1}$ &                \\
    \hline
    \multicolumn{4}{c}{pulsar}  \\
    $0.34^{+0.16}_{-0.16}$ & $1.46^{+0.34}_{-0.35}$ &  $3.6^{+1.8}_{-1.3}$                        & 245 (649)  \\   
    \multicolumn{4}{c}{pulsar+PWN}  \\
    $0.60^{+0.18}_{-0.15}$ & $1.59^{+0.31}_{-0.28}$ &  $12.8^{+5.3}_{-3.5}$                       & 448 (649)  \\  
    \multicolumn{4}{c}{south-east tail of the PWN}  \\
    $1.90^{+0.71}_{-0.61}$ & $2.11^{+0.69}_{-0.61}$ &  $11.8^{+17.1}_{-6.6}$                      & 267 (649)  \\  
    \hline 
  \end{tabular}
  \label{t:x-fit}
\end{table}

There is  
a noticeable increase of the  absorbing column 
density $N_H$ when we move from the
point-like pulsar to the extended PWN.
This increase is 
significant for the south-east tail,  
which overlaps 
with the optically dark region 
in Fig.~\ref{fig:pulsar_vicinity}  
and, therefore, 
is most strongly absorbed also in X-rays.   
The pulsar and  north-western part of the PWN are   
in a more transparent region 
and  $N_H$  is   
lower for the pulsar and has an intermediate value for the entire
system. We therefore believe that the $N_H$
derived from the 1\farcs5  aperture
is more realistic for the pulsar 
than the factor of 2 higher value obtained from the analysis of the
pulsar+PWN spectrum.     

\subsection{The interstellar extinction}
\label{sec:extinction}
The next step is dereddening the optical data.   
$N_H$ values from 
Table~\ref{t:x-fit} and a standard
relation between 
$N_H$ and the extinction factor $A_{V}$ 
 \citep{predehl1995A&A},  yield
$A_{V}$ of  3.4$^{+1.0}_{-0.8}$, 
10.6$^{+4.0}_{-3.4}$, and 1.8$^{+0.9}_{-0.9}$ 
for the pulsar+PWN, PWN tail, and
pulsar, respectively. The first
value is nearly equal to 
the entire Galactic extinction of      
3.6 mag along the pulsar line-of-sight 
\citep{schlegel1998ApJ}. The
second one is significantly larger,
in agreement  with the absence of
any stars  in the dark region overlapping with
the tail.    
The pulsar itself is apparently less reddened.   
To verify that,   
we made independent $A_{V}$
estimates   using 
a method based on the red-clump stars as standard candles,
which provides an $A_{V}$--distance
relation for a given position on the sky \citep[see e.g.][]{lopez2002A&A}.
\begin{figure}
  \setlength{\unitlength}{1mm}
  \resizebox{15.5cm}{!}{   
  \begin{picture}(50,16)(0,0)
    \put (5,0) {\includegraphics[scale=0.080]{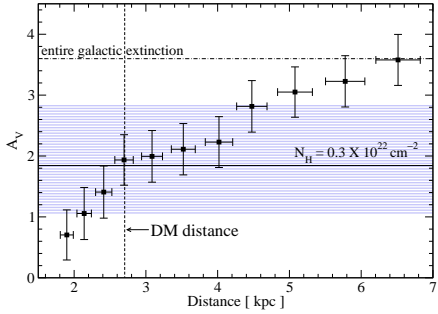}}
  \end{picture}
  }
  \caption{
  The $A_{V}$--distance relation towards
  \psr. The vertical dashed line corresponds to the DM distance to the pulsar of  
  2.7 kpc. The hatched region  and horizontal solid line show the 
  confidence interval of $A_{V}$ resulting from  the uncertainty
  interval of    
  $N_H$  and  its  best-fit value 
  for the pulsar presented in Table~\ref{t:x-fit},
  respectively.   
  The entire galactic extinction  
  is indicated by 
  the dash-dotted line.} 
  \label{fig:red-clump}
\end{figure}

We extracted stars 
located within 0.3\degs~of the pulsar position 
from the 2MASS All-Sky Point Source 
Catalogue\footnote{See http://irsa.ipac.caltech.edu/applications/DataTag/, 
DataTag $=$ ADS/IRSA.Gator\#2012/0306/095311\_11593.},
and created  a  colour-magnitude diagram,  
$K$ {\sl vs} $J-K$. 
We found  mean $J-K$ colours of the
red-clump branch in  several 
magnitude bins     
and  transformed  them  to the
$A_{V}$--distance relation,    
as has been done by  
\citet{danilenko2012A&A} for
another $\gamma$-ray
pulsar J1357$-$6429. 
At a large distance limit 
this relation (Fig.~\ref{fig:red-clump}) is in agreement 
with the entire Galactic
extinction in this direction. 
For the DM  pulsar distance of 2.7 kpc (the
vertical dashed line  in
Fig.~\ref{fig:red-clump})   
it suggests $A_{V}$~$\approx$ 2, which is
consistent  with   
the  X-ray spectral fit 
(the horizontal solid line) 
and with a
mean foreground $A_{V}$~$\approx$~1.7 for the Carina
complex \citep{hillier2001ApJ}.  

Within the  dark regions  
discussed above   
$A_V$ is obviously larger
\citep[cf.][]{povich2011ApJS}, but these    
have a negligible contribution to   
the derived  relation  dominated  
by more transparent  regions, 
in one of which the pulsar is located.

We thus consider $A_{V}$~$\approx$ 1.8,    
$N_H$~$\approx$ 3~$\times$~10$^{21}$ cm$^{-2}$, and d~$\approx$ 2.7 kpc 
as the most appropriate
values, and hereafter use them for
compilation of  multi-wavelength spectrum and
luminosity estimates of \psr.  
\begin{figure}[t]
  \setlength{\unitlength}{1mm}
  \resizebox{15.5cm}{!}{   
  \begin{picture}(50,17)(0,0)
    \put (5,0)  {\includegraphics[scale=0.085]{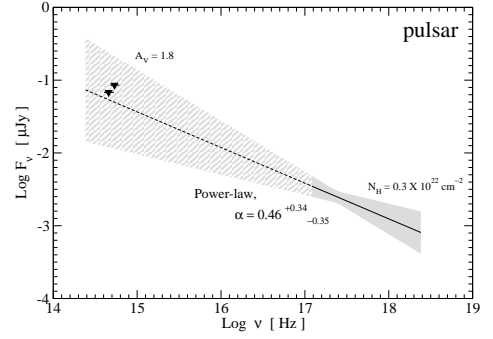}}
  \end{picture}
  }
  \caption{Unabsorbed
  multi-wavelength spectrum of \psr~compiled  
  from data obtained with the \textit{Chandra} 
  and VLT. The X-ray part with its
  uncertainties (hatched regions) is extrapolated to the
  optical.  Black triangles 
  are  optical  flux   upper
  limits.
  }
  \label{fig:multi_spec}
\end{figure}

\subsection{Multi-wavelength spectrum and luminosities}
The unabsorbed multi-wavelength spectrum of the pulsar 
is shown in Fig.~\ref{fig:multi_spec}. Its X-ray part 
is obtained with $N_H$ frozen at the value of 
3~$\times$ 10$^{21}$ cm$^{-2}$, and the optical  data 
are dereddened with $A_{V}$~= 1.8.  
Our upper limits show that the real optical fluxes of the pulsar 
cannot significantly exceed the extrapolation of its X-ray spectrum  
to the optical range. This is typical  for  non-thermal radiation  
from pulsars  detected in the optical and  X-rays, including  
the Crab \citep{sandberg2009A&A}, Vela  \citep{shibanov2003A&A},  some
middle-aged \citep{shibanov2006A&A}, and old  pulsars 
\citep{zharikov2004A&A,zavlin2004ApJ}.
 
\begin{figure*}[t]
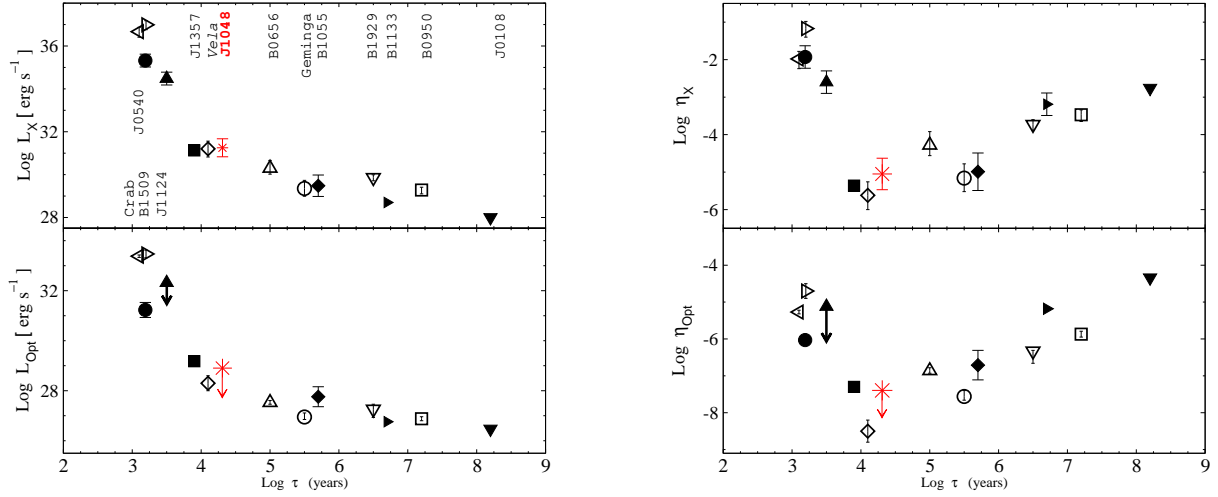

  \begin{center}
  \setlength{\unitlength}{1mm}
  \resizebox{15.5cm}{!}{   
  \begin{picture}(50,23)(0,0)
    \put (-2,0)
    {\includegraphics[scale=0.115]{AgeLumOpX_1.eps}}
    \put (27,0)
    {\includegraphics[scale=0.115]{AgeLumOpX.eps}}
  \end{picture}
  }
  \end{center}
  \caption{Comparison of  X-ray and $V$-band luminosities and efficiencies 
  for pulsars of different characteristic age $ \tau$ 
  detected in both spectral domains. Different pulsars are marked 
  by different symbols and \psr~is shown by the red star. 
  Both dependencies of  pulsar 
  efficiencies with age  demonstrate an efficiency minimum near the Vela age,  
  and the Vela-like J1048$-$5832 naturally occupies a place near this minimum.   
  }
  \label{fig:opt_lum}
\end{figure*}

Using our spectral fits, the pulsar X-ray
luminosity  in the 2--10 keV range  at 
d~= 2.7 kpc, 
$L_{X}^{psr}$~$\approx$ 10$^{31.3}$ erg~s$^{-1}$,   
and the pulsar to pulsar+PWN X-ray luminosity  
ratio  $\approx$~0.46  are very similar to those 
of $\approx$~10$^{31.2}$ erg~s$^{-1}$ and 0.34,   
respectively, obtained for the Vela pulsar,   
which has a similar age and spin-down luminosity. 
Comparing this together with our constraints
of optical luminosity 
$L_{opt}$~$\la$ 10$^{28.9}$ erg~s$^{-1}$ and  the efficiency of
transformation of its spin-down luminosity $\dot{E}$ to  optical
photons $L_{opt}/\dot{E}$~$\la$ 10$^{-7.4}$ 
with the data available 
for other pulsars detected in the
optical and X-rays
\citep{zharikov2004A&A,danilenko2012A&A},  
we conclude that, as the Vela pulsar, \psr~is  rather  inefficient 
in the optical and X-rays (Fig.~\ref{fig:opt_lum}). 

\section{Discussion and conclusions}
\label{sec4}  
We did not detect  \psr~in our deep 
optical images of the  field,  down
to a visual magnitude  of  
$\sim$~28. 
The pulsar is located in a complicated region
linked to the northern edge of the Carina complex with   
high star formation (and supernova)
activity (Fig.~\ref{fig:1}). The region 
is filled by clumpy 
clouds where the interstellar
extinction and  absorbing column density
vary substantially even at a
10\asec~scale.  
This complicates
the optical identification of the pulsar and  its PWN. 
Nevertheless, the derived pulsar optical
flux upper limits are quite informative.

First, our VLT observations and reanalysis of the X-ray
data  show (Fig.~\ref{fig:multi_spec}) that   the optical fluxes
of the pulsar do not exceed the  extrapolation of its power law 
X-ray spectrum towards the optical range,   which is compatible 
with multi-wavelength non-thermal spectra for
other rotation powered pulsars. 

Second,  our results show that this Vela-like 
pulsar is very 
inefficient in the optical and X-rays, 
as is the Vela pulsar itself (Fig.~\ref{fig:opt_lum}). 
In combination  with a significant interstellar extinction 
towards the pulsar, $A_{V}$~$\approx$ 2, this precluded us to
detect it in our 
deep observations at a level consistent with the optical 
efficiency of the Vela pulsar.   

We did detect a faint optical source  coinciding with a relatively 
bright compact feature of  the pulsar
X-ray PWN, presumably a wisp, located 
$\sim$~2\asec~from the pulsar
(Fig.~\ref{fig:pulsar_vicinity2}). 
The source colour $V-R$~$\approx$ 0.6 
is consistent with colours typical for PWNe structures,  
suggesting that the source is an optical 
counterpart candidate to the X-ray feature. 
The poor X-ray count
statistics precludes us to constrain 
the multi-wavelength spectrum
of the source and its background nature
cannot be excluded. 

Any further optical
studies of the pulsar  have to be
carried out  at longer wavelengths, which are less affected
by the interstellar 
extinction.

Recent progress in the multi-wavelength studies of Vela-like 
pulsars adds new  
evidence  of their low efficiency in the optical and X-rays.    
This forms a puzzling minimum in the optical
and X-ray efficiency  relations 
 {\sl vs} pulsar age, 
while in  $\gamma$-rays it appears to be absent \citep{abdo2010ApJS_Fermi_Catalog}.  
Further studies  will show  
whether this is a signature of some interesting changes in a neutron star  
magnetosphere  and particle acceleration at the 10 kyr age, 
or just an incomplete sample effect, which disappear when more Vela-like 
pulsars  will be detected in the optical and X-rays. 

\begin{acknowledgements} 
We are grateful to anonymous referee for useful 
comments allowing us to improve the paper. 
The work was partially supported by the Russian Foundation for Basic 
Research (grants 11-02-00253 and 11-02-12082),
RF Presidential Program (Grant NSh 4035.2012.2), and
the Ministry of  Education and Science of the Russian Federation 
(Contract No. 11.G34.31.0001 and Agreement No.8409, 2012). 
\end{acknowledgements}

{\small \noindent Note added: After submission of this paper also Razzano et al. (2013, MNRAS, 428, 3636)
appeared. It is based on our data and their conclusion that star 6 cannot be the
optical counterpart to the pulsar is consistent with our results while their pulsar flux
upper limits are less deeper.}

\bibliographystyle{aa}
\bibliography{ref-1}
\end{document}